\documentstyle[prl,eqsecnum,aps]{revtex}
\begin{document}
\author{Jian-Qi Shen $^{1,}$$^{2}$ \footnote{E-mail address: jqshen@coer.zju.edu.cn} and Zhi-Chao Ruan $^{1}$}
\address{$^{1}$  Centre for Optical
and Electromagnetic Research, State Key Laboratory of Modern
Optical Instrumentation, \\Zhejiang University,
Hangzhou Yuquan 310027, P.R. China\\
$^{2}$ Zhejiang Institute of Modern Physics and Department of
Physics, Zhejiang University, Hangzhou 310027, P.R. China}
\date{\today }
\title{Maintenance of Coherence and Polarization Evolution \\in a Supersymmetric Multiphoton Model}
\maketitle

\begin{abstract}
The elimination of decoherence of two-state quantum systems
interacting with a thermal reservoir through an external
controllable driving field is discussed in the present paper. The
restriction equation with which the external controllable driving
field should agree will be derived. Based on this, we obtain the
time-development equation of the off-diagonal elements of density
operator in the supersymmetric multiphoton two-state quantum
systems, which is helpful for studying the polarization evolution
in this two-state quantum model.


\end{abstract}

\section{Introduction}
Recently, an area called {\it quantum computation}, which involves
computers that use the ability of quantum systems to be in a
superposition of many states, attracts extensive attention of many
researchers. However, it is not yet clear whether quantum
computers are feasible to build\cite{Shor}. One reason that
quantum computers will be difficult, if not impossible, to build
is {\it decoherence}. In the process of decoherence, some qubit or
qubits of the computation become entangled with the environment,
thus in effect ``collapsing'' the state of the quantum
computer\cite{Shor}. In literature, there may exist three schemes
to reduce the decoherence: (i) error-avoiding codes\cite{Zanardi};
(ii) error-correcting codes\cite{Gottesman}; (iii)
decoherence-avoiding scheme\cite{Viola}. The third approach to the
suppression of decoherence can be realized by eliminating the
interaction between the (two-state) quantum system and environment
(such as a noise field, bath, thermal reservoir and so on) in the
presence of an external controllable driving field. In this paper,
we will study the maintenance of coherence via the
decoherence-avoiding scheme, and the time evolution of
polarization in a supersymmetric multiphoton two-level model. On
considering the latter problem, we assume that the environmental
effect on the quantum system under consideration has been
eliminated by using the decoherence-avoiding scheme, which enables
us to treat the polarization evolution problem in the multiphoton
model more conveniently ({\it i.e.}, the polarization evolution
problem will be investigated under the assumption that the
decoherence of the quantum systems in the noise field has been
reduced).

In section II, we introduce Zhang's method for treating the
decoherence problem\cite{Zhang}. In section III, by employing
Zhang's approach to the multiphoton two-state quantum system we
obtain the time-development equation of the off-diagonal element
of density operator in this two-state system.

\section{Maintenance of Coherence}
In this section, we review one of the formulation for dealing with
the decoherence-avoiding scheme, which was suggested by
Zhang\cite{Zhang} more recently. The reason for the detailed
reappearance of Zhang\cite{Zhang} below is as follows: (i) first
and foremost, Zhang's formulation is helpful for treating the
polarization evolution problem in the multiphoton model; (ii) in
this paper, the elimination of decoherence is a prerequisite for
simplifying the polarization evolution problem in the multiphoton
model, namely, in studying the polarization evolution problem in
the multiphoton two-state system, we assume that the entanglement
of the quantum system with the environment ({\it e.g.}, thermal
reservoir) has been eliminated by an external driving field. So,
we need not consider the decoherence problem of the multiphoton
system in section III.

Let us first consider the following model, the Hamiltonian of
which takes the form (in the unit $\hbar=1$)

\begin{equation}
H(t)=\frac{\omega_{0}}{2}\sigma_{z}+\sum_{k}\omega_{k}a^{\dagger}_{k}a_{k}+\sum_{k}g_{k}\sigma_{z}
(a_{k}+a^{\dagger}_{k})-\frac{d}{2}[E(t)\sigma_{+}+E^{\ast}(t)\sigma_{-}],
\label{eq21}
\end{equation}
which can describe the interaction of a two-level atom with a
noise field (thermal reservoir). In this Hamiltonian,
$\omega_{0}$, $\omega_{k}$, $\sigma_{z}$, $g_{k}$ and $E(t)$
denote the atomic transition frequency, photon frequency with $k$-
mode, the third-component Pauli matrix, the coupling coefficient
(of atoms to the thermal reservoir) and the external driving
field, respectively. By using the unitary transformation
\begin{equation}
V(t)=\exp\left[\frac{1}{i}\left(\frac{\omega_{0}}{2}\sigma_{z}+\sum_{k}\omega_{k}a^{\dagger}_{k}a_{k}\right)t\right],
\end{equation}
one can arrive at the Hamiltonian
\begin{equation}
H_{\rm I}(t)=\sum_{k}g_{k}\sigma_{z}
\left[a_{k}\exp(-i\omega_{k}t)+a^{\dagger}_{k}\exp(i\omega_{k}t)\right]-\frac{d}{2}\left[E(t)\exp(i\omega_{0}t)\sigma_{+}+E^{\ast}(t)\exp(-i\omega_{0}t)\sigma_{-}\right]
\end{equation}
in the interaction picture, where use is made of $H_{\rm
I}(t)=V^{\dagger}(t)\left[H(t)-i\frac{\partial}{\partial t
}\right]V(t)$. The density operator of the two-level atomic system
agrees with
\begin{equation}
i\frac{\partial \rho_{\rm I}(t)}{\partial t}=[H_{\rm I}(t),
\rho_{\rm I}(t)].
\end{equation}
Let $\rho_{q\rm I}(t)$ denote the atomic reducible density
operator. It follows that the reducible density operator equals
\begin{equation}
\dot{\rho}_{q\rm I}(t)={\rm Tr}_{\rm r}\dot{\rho}_{\rm
I}(t)=-\int^{t}_{0}{\rm Tr}_{\rm r}[H_{\rm I}(t), [H_{\rm I}(t'),
\rho_{\rm I}(t')]]{\rm d}t'.                   \label{eq25}
\end{equation}
If we assume that the thermal reservoir is rather large, then it
can be concluded that the reservoir may not change much during the
time evolution process of the atom-reservoir system. Thus we have
$\rho_{\rm I}(t)\simeq {\rho}_{q\rm I}(t){\rho}_{{\rm r}{\rm
I}}(0)$, where ${\rho}_{{\rm r}{\rm I}}(0)\simeq \exp (-\beta
H_{0})/{\rm Tr}\exp(-\beta H_{0})$ with $\beta=1/k_{\rm B}T$,
$H_{0}=\sum_{k}\omega_{k}a^{\dagger}_{k}a_{k}$. If we take
$\rho_{\rm I}(t')\simeq \rho_{\rm I}(t)$ (Markoff approximation),
then Eq.(\ref{eq25}) can be rewritten as
\begin{equation}
\dot{\rho}_{q\rm I}(t)=-\int^{t}_{0}{\rm Tr}_{\rm r}[H_{\rm I}(t),
[H_{\rm I}(t'), {\rho}_{q\rm I}(t){\rho}_{{\rm r}{\rm I}}(0)]]{\rm
d}t',                              \label{eq26}
\end{equation}
where the integrand can be rewritten as
\begin{eqnarray}
{\rm Tr}_{\rm r}[H_{\rm I}(t), [H_{\rm I}(t'), {\rho}_{q\rm
I}(t){\rho}_{{\rm r}{\rm I}}(0)]]&=&{\rm Tr}_{\rm r}[H_{\rm
I}(t)H_{\rm I}(t'){\rho}_{q\rm I}(t){\rho}_{{\rm r}{\rm
I}}(0)-H_{\rm
I}(t){\rho}_{q\rm I}(t){\rho}_{{\rm r}{\rm I}}(0)H_{\rm I}(t')                  \nonumber \\
&-&H_{\rm I}(t'){\rho}_{q\rm I}(t){\rho}_{{\rm r}{\rm I}}(0)H_{\rm
I}(t)+{\rho}_{q\rm I}(t){\rho}_{{\rm r}{\rm I}}(0)H_{\rm
I}(t')H_{\rm I}(t)].     \label{eq27}
\end{eqnarray}
By the aid of the relations $<a^{\dagger}_{k}a_{k}>_{\rm
R}=\bar{n}_{k}$, $<a_{k}a^{\dagger}_{k}>_{\rm R}=\bar{n}_{k}+1$,
$\bar{n}_{k}=1/[\exp(\hbar \omega_{k}/k_{\rm B}T)-1]$, one can
arrive at through lengthy calculation\cite{Zhang}
\begin{equation}
\dot{\rho}_{q\rm
I}(t)=-2\left[\left(-\frac{d}{2}\right)^{2}{\rho}_{q\rm
I}(t)|E(t)|^{2}\int^{t}_{0}\exp[i\omega_{0}(t-t')]{\rm
d}t'-{\rho}_{q\rm I}(t){\mathcal A}+\sigma_{z}{\rho}_{q\rm
I}(t)\sigma_{z}{\mathcal A}\right]
\end{equation}
with ${\mathcal A}=A_{1}+A_{2}+A_{3}+A_{4}$, where
\begin{eqnarray}
A_{1}&=&\sum_{k}g_{k}^{2}(\bar{n}_{k}+1)\frac{1-\exp(-i\omega_{k}t)}{i\omega_{k}},               \quad         A_{2}=\sum_{k}g_{k}^{2}\bar{n}_{k}\frac{1-\exp(i\omega_{k}t)}{-i\omega_{k}},         \nonumber \\
A_{3}&=&\sum_{k}g_{k}^{2}(\bar{n}_{k}+1)\frac{1-\exp(i\omega_{k}t)}{-i\omega_{k}},
\quad
A_{4}=\sum_{k}g_{k}^{2}\bar{n}_{k}\frac{1-\exp(-i\omega_{k}t)}{i\omega_{k}}.
\end{eqnarray}
Ignoring the Lamb-shift term ({\it i.e.}, taking the real parts of
$\dot{\rho}_{01}=\langle 0|\dot{\rho}_{q \rm I}|1\rangle$,
${\rho}_{01}=\langle 0|{\rho}_{q \rm I}|1\rangle$,
$\dot{\rho}_{10}=\langle 1|\dot{\rho}_{q \rm I}|0\rangle$,
${\rho}_{10}=\langle 1|{\rho}_{q \rm I}|0\rangle$), we can obtain
the time-development equation of the off-diagonal elements of
density operator, {\it i.e.},

\begin{eqnarray}
\dot{\rho}_{01}=-2\left[\frac{\sin (\omega_{0}t)}{\omega_{0}}\frac{d^{2}}{4}|E(t)|^{2}
+\sum_{k}2g_{k}^{2}(2\bar{n}_{k}+1)\frac{\sin (\omega_{k}t)}{\omega_{k}}\right]{\rho}_{01},                 \nonumber \\
\dot{\rho}_{10}=-2\left[\frac{\sin
(\omega_{0}t)}{\omega_{0}}\frac{d^{2}}{4}|E(t)|^{2}+\sum_{k}2g_{k}^{2}(2\bar{n}_{k}+1)\frac{\sin
(\omega_{k}t)}{\omega_{k}}\right]{\rho}_{10}.
\end{eqnarray}
Thus it is readily verified that if the envelope of the external
driving field satisfies the following condition

\begin{equation}
\frac{\sin
(\omega_{0}t)}{\omega_{0}}\frac{d^{2}}{4}|E(t)|^{2}+\sum_{k}2g_{k}^{2}(2\bar{n}_{k}+1)\frac{\sin
(\omega_{k}t)}{\omega_{k}}=0,
\end{equation}
then one have $\rho_{01}(t)=\rho_{01}(0)$ and
$\rho_{10}(t)=\rho_{10}(0)$, which means the suppression
(elimination) of decoherence in this two-level quantum system
interacting with the environment (thermal reservoir) through an
external controllable field $E(t)$.

It should be noted again that the above theory was proposed by
Zhang\cite{Zhang}. In the next section we will investigate the
polarization evolution problem in the two-level supersymmetric
multiphoton Jaynes-Cummings model by making use of Zhang's
formulation.

\section{Polarization evolution in the multiphoton two-level quantum system}
The multiphoton two-level system that we will consider in this
section is the supersymmetric multiphoton Jaynes-Cummings
model\cite{Sukumar,Kien}, the Hamiltonian of which under the
rotating wave approximation is given by

\begin{equation}
H=\frac{\omega _{0}}{2}\sigma _{z}+\omega a^{\dagger
}a+g(a^{\dagger })^{k}\sigma _{-}+g^{\ast }a^{k}\sigma _{+},
\label{eq31}
\end{equation}
where $a^{\dagger }$ and $a$ are the creation and annihilation
operators for the electromagnetic field, and obey the commutation
relation $\left[ a,a^{\dagger }\right] =1$; $\sigma _{\pm }$ and
$\sigma _{z}$ denote the two-level atom operators which satisfy
the commutation relation $\left[
\sigma _{z},\sigma _{\pm }\right] =\pm 2\sigma _{\pm }$ ; $g(t)$ and $%
g^{\ast }(t)$ are the coupling coefficients and $k$ is the photon
number in each atom transition process; $\omega _{0}(t)$ and
$\omega (t)$ are respectively the transition frequency and the
mode frequency.

The supersymmetric structure can be found in this multiphoton
two-level quantum model by defining the following supersymmetric
transformation generators\cite{Lu1,Lu2}:

\begin{eqnarray}
N &=&a^{\dagger }a+\frac{k-1}{2}\sigma _{z}+\frac{1}{2}=\left(
\begin{array}{cc}
a^{\dagger }a+\frac{k}{2} & 0 \\
0 & aa^{\dagger }-\frac{k}{2}
\end{array}
\right) ,\quad N^{^{\prime }}=\left(
\begin{array}{cc}
a^{k}(a^{\dagger })^{k} & 0 \\
0 & (a^{\dagger })^{k}a^{k}
\end{array}
\right) ,  \nonumber \\
Q &=&(a^{\dagger })^{k}\sigma _{-}=\left(
\begin{array}{cc}
0 & 0 \\
(a^{\dagger })^{k} & 0
\end{array}
\right) ,\quad Q^{\dagger }=a^{k}\sigma _{+}=\left(
\begin{array}{cc}
0 & a^{k} \\
0 & 0
\end{array}
\right) .  \label{eq32}
\end{eqnarray}
It is easily verified that $(N,N^{^{\prime }},Q,Q^{\dagger })$
form supersymmetric generators and have supersymmetric Lie algebra
properties, {\it i.e.},

\begin{eqnarray}
Q^{2} &=&(Q^{\dagger })^{2}=0,\quad \left[ Q^{\dagger },Q\right]
=N^{^{\prime }}\sigma _{z},\quad \left[ N,N^{^{\prime }}\right] =0,\quad %
\left[ N,Q\right] =Q,  \nonumber \\
\left[ N,Q^{\dagger }\right] &=&-Q^{\dagger },\quad \left\{
Q^{\dagger },Q\right\} =N^{^{\prime }},\quad \left\{ Q,\sigma
_{z}\right\} =\left\{
Q^{\dagger },\sigma _{z}\right\} =0,  \nonumber \\
\left[ Q,\sigma _{z}\right] &=&2Q,\quad \left[ Q^{\dagger },\sigma _{z}%
\right] =-2Q^{\dagger },\quad \left( Q^{\dagger }-Q\right)
^{2}=-N^{^{\prime }},        \label{eq33}
\end{eqnarray}
where $\left\{ {}\right\} $ denotes the anticommuting bracket.

Now let us obtain the Hamiltonian of the above multiphoton
Jaynes-Cummings model in the interaction picture by using the
following unitary transformation
\begin{equation}
V(t)=\exp\left[\frac{1}{i}\left(\frac{\omega _{0}}{2}\sigma
_{z}+\omega a^{\dagger }a\right)t\right],
\end{equation}
and the result is
\begin{equation}
H_{\rm I}(t)=g\exp(-i\delta t)Q+g^{\ast}\exp(i\delta t)Q^{\dagger}
\end{equation}
with $\delta=k\omega-\omega _{0}$. Based on Eq.(\ref{eq27}), by
complicated calculation, one can arrive at
\begin{eqnarray}
{\rm Tr}_{\rm r}[H_{\rm I}(t)H_{\rm I}(t'){\rho}_{q\rm
I}(t){\rho}_{{\rm r}{\rm I}}(0)]&=&<H_{\rm I}(t)H_{\rm
I}(t'){\rho}_{q\rm I}(t)>_{\rm R}        \nonumber \\
&=&gg^{\ast}\left\{\exp[i\delta(t'-t)]<QQ^{\dagger}\rho_{q\rm
I}(t)>_{\rm R} +\exp[-i\delta(t'-t)]<Q^{\dagger}Q\rho_{q\rm
I}(t)>_{\rm R}\right\},             \label{eq271}
\end{eqnarray}
where $Q^{2}=(Q^{\dagger})^{2}=0$ is applied to the calculation,
and
\begin{eqnarray}
{\rm Tr}_{\rm r}[H_{\rm I}(t){\rho}_{q\rm I}(t){\rho}_{{\rm r}{\rm
I}}(0)H_{\rm I}(t')]&=&<H_{\rm I}(t){\rho}_{q\rm I}(t)H_{\rm
I}(t')>_{\rm R}                                                      \nonumber \\
&=&g^{2}\exp[-i\delta (t+t')]<Q{\rho}_{q\rm I}(t)Q>_{\rm R}+(g^{\ast})^{2}\exp[i\delta (t+t')]<Q^{\dagger}{\rho}_{q\rm I}(t)Q^{\dagger}>_{\rm R}      \nonumber \\
&+&gg^{\ast}\left[\exp[i\delta(t'-t)]<Q{\rho}_{q\rm
I}(t)Q^{\dagger}>_{\rm
R}+\exp[-i\delta(t'-t)]<Q^{\dagger}{\rho}_{q\rm I}(t)Q>_{\rm
R}\right],
\end{eqnarray}

\begin{eqnarray}
{\rm Tr}_{\rm r}[H_{\rm I}(t'){\rho}_{q\rm I}(t){\rho}_{{\rm
r}{\rm I}}(0)H_{\rm I}(t)]&=&g^{2}\exp[-i\delta (t+t')]<Q{\rho}_{q\rm I}(t)Q>_{\rm R}+(g^{\ast})^{2}\exp[i\delta (t+t')]<Q^{\dagger}{\rho}_{q\rm I}(t)Q^{\dagger}>_{\rm R}      \nonumber \\
&+&gg^{\ast}\left[\exp[-i\delta(t'-t)]<Q{\rho}_{q\rm
I}(t)Q^{\dagger}>_{\rm
R}+\exp[i\delta(t'-t)]<Q^{\dagger}{\rho}_{q\rm I}(t)Q>_{\rm
R}\right],
\end{eqnarray}
and
\begin{eqnarray}
{\rm Tr}_{\rm r}[{\rho}_{q\rm I}(t){\rho}_{{\rm r}{\rm
I}}(0)H_{\rm I}(t')H_{\rm
I}(t)]&=&gg^{\ast}\left\{\exp[-i\delta(t'-t)]<\rho_{q\rm
I}(t)QQ^{\dagger}>_{\rm R} +\exp[i\delta(t'-t)]<\rho_{q\rm
I}(t)Q^{\dagger}Q>_{\rm R}\right\}.            \label{eq274}
\end{eqnarray}
Thus it follows from (\ref{eq27}) and the above four expressions
(\ref{eq271})-(\ref{eq274}) that
\begin{equation}
{\rm Tr}_{\rm r}[H_{\rm I}(t), [H_{\rm I}(t'), {\rho}_{q\rm
I}(t){\rho}_{{\rm r}{\rm I}}(0)]]={\mathcal T}_{1}+{\mathcal
T}_{2}+{\mathcal T}_{3},
\end{equation}
where
\begin{eqnarray}
{\mathcal
T}_{1}&=&gg^{\ast}\left\{\exp[i\delta(t'-t)]<(a^{\dagger})^{k}a^{k}>_{\rm
R}\sigma_{-}\sigma_{+}+\exp[-i\delta(t'-t)]<a^{k}(a^{\dagger})^{k}>_{\rm
R}\sigma_{+}\sigma_{-}\right\}{\rho}_{q\rm
I}(t)                           \nonumber\\
&+&gg^{\ast}{\rho}_{q\rm
I}(t)\left\{\exp[-i\delta(t'-t)]<(a^{\dagger})^{k}a^{k}>_{\rm
R}\sigma_{-}\sigma_{+}+\exp[i\delta(t'-t)]<a^{k}(a^{\dagger})^{k}>_{\rm
R}\sigma_{+}\sigma_{-}\right\},
\end{eqnarray}

\begin{equation}
{\mathcal T}_{2}=-2\left\{g^{2}\exp[-i\delta (t+t')]<Q{\rho}_{q\rm
I}(t)Q>_{\rm R}+(g^{\ast})^{2}\exp[i\delta
(t+t')]<Q^{\dagger}{\rho}_{q\rm I}(t)Q^{\dagger}>_{\rm R}\right\}
\end{equation}
and
\begin{equation}
{\mathcal T}_{3}=-gg^{\ast}\left\{\exp[i\delta
(t'-t)]+\exp[-i\delta (t'-t)]\right\}\left(<Q{\rho}_{q\rm
I}(t)Q^{\dagger}>_{\rm R}+<Q^{\dagger}{\rho}_{q\rm I}(t)Q>_{\rm
R}\right).
\end{equation}
With the help of the following relations $a^{k}(a^{\dagger
})^{k}\left| m\right\rangle =\frac{(m+k)!}{ m!}\left|
m\right\rangle $, $<a^{k}(a^{\dagger})^{k}>_{\rm R}=\frac{(m+k)!}{
m!}$, $<(a^{\dagger})^{k}a^{k}>_{\rm R}=\frac{m!}{ (m-k)!}$,
$<(a^{\dagger})^{k}(a^{\dagger})^{k}>_{\rm R}=0$,
$\sigma_{+}|+\rangle=0$,  $\sigma_{-}|-\rangle=0$,
$\langle+|\sigma_{-}=0 $,   $\langle-|\sigma_{+}=0 $, one can
arrive at
\begin{eqnarray}
\langle0|{\mathcal
T}_{1}|1\rangle&=&gg^{\ast}\exp[i\delta(t'-t)]\left[\frac{(m+k)!}{
m!}+\frac{m!}{(m-k)!}\right]\langle0|{\rho}_{q\rm I}(t)|1\rangle,
 \nonumber   \\
\langle0|{\mathcal T}_{2}|1\rangle&=&0,
\nonumber   \\
 \langle 0|{\mathcal
T}_{3}|1\rangle&=&-gg^{\ast}\left\{\exp[i\delta
(t'-t)]+\exp[-i\delta (t'-t)]\right\}\frac{(m+k)!}{m!}\langle
1|{\rho}_{q\rm I}(t)|0\rangle.
\end{eqnarray}
If we set ${\rho}_{01}(t)=\langle 0|{\rho}_{q\rm I}(t)|1\rangle$,
${\rho}_{10}(t)=\langle 1|{\rho}_{q\rm I}(t)|0\rangle$,
$\dot{{\rho}}_{01}(t)=\langle 0|\dot{{\rho}}_{q\rm
I}(t)|1\rangle$, and use the equation
\begin{eqnarray}
\dot{{\rho}}_{01}(t)=-\int^{t}_{0}{\rm Tr}_{\rm r}\langle
0|[H_{\rm I}(t), [H_{\rm I}(t'), {\rho}_{q\rm I}(t){\rho}_{{\rm
r}{\rm I}}(0)]]|1\rangle{\rm d}t'          \label{eq3160}
\end{eqnarray}
in accordance with (\ref{eq26}), then we will get
\begin{equation}
\dot{{\rho}}_{01}(t)=c_{1}(t){\rho}_{01}(t)-c_{2}(t){\rho}_{10}(t)
\label{eq316}
\end{equation}
with
\begin{equation}
c_{1}(t)=-gg^{\ast}\left[\frac{(m+k)!}{
m!}+\frac{m!}{(m-k)!}\right]\frac{1-\exp(-i\delta t )}{i\delta},
\quad
c_{2}(t)=-gg^{\ast}\frac{(m+k)!}{m!}\left[\frac{1-\exp(-i\delta t
)}{i\delta}+\frac{1-\exp(i\delta t )}{-i\delta}\right].
\label{eqeq}
\end{equation}
In the meanwhile, we can obtain the complex conjugation to
Eq.(\ref{eq316}), {\it i.e.},
\begin{equation}
\dot{{\rho}}_{10}(t)=c^{\ast}_{1}(t){\rho}_{10}(t)-c_{2}(t){\rho}_{01}(t).
\label{eq317}
\end{equation}
Thus we obtain the time-development equation of the off-diagonal
elements of density operator in this supersymmetric multiphoton
two-state quantum systems.

In order to indicate the physical meanings of Eqs.(\ref{eq316})
and (\ref{eq317}), we will set
 $u=\rho_{01}+\rho_{10}$,   $v=i(\rho_{01}-\rho_{10})$. Eqs.(\ref{eq316})
and (\ref{eq317}) can therefore be rewritten as follows
\begin{eqnarray}
\frac{{\rm d}u}{{\rm d}t}&=&\left[{\rm Re}c_{1}(t)-c_{2}(t)\right]u+{\rm Im}c_{1}(t)v,                  \nonumber \\
\frac{{\rm d}v}{{\rm d}t}&=&\left[{\rm
Re}c_{1}(t)+c_{2}(t)\right]v-{\rm Im}c_{1}(t)u. \label{eq318}
\end{eqnarray}
It is easily seen that here the physical meanings of $u$ and $v$
are as follows: $u$ and $v$ have close relation to the dispersion
and the dissipation (absorption/amplification) of polarization,
respectively.
\section{Brief discussion}
To close this section, we briefly discuss the time-evolution
equation (\ref{eq318}) of $u$ and $v$. Let us consider a simple
case where the detuning frequency $\delta$ from the atomic
transition frequency is vanishing. It follows from the expressions
(\ref{eqeq}) that $c_{1}(t)=-gg^{\ast}\left[\frac{(m+k)!}{
m!}+\frac{m!}{(m-k)!}\right]t$ and
$c_{2}(t)=-2gg^{\ast}\frac{(m+k)!}{m!}t$. Note that here
$c_{1}(t)$ is a real function. So, according to Eq.(\ref{eq318}),
we obtain
\begin{equation}
u(t)=u_{0}\exp\left[\frac{1}{2}\lambda_{u}t^{2}\right],  \qquad
v(t)=v_{0}\exp\left[\frac{1}{2}\lambda_{v}t^{2}\right],
\end{equation}
where
$\lambda_{u}=-gg^{\ast}\left[\frac{m!}{(m-k)!}-\frac{(m+k)!}{
m!}\right]$ and
$\lambda_{v}=-gg^{\ast}\left[\frac{m!}{(m-k)!}+3\frac{(m+k)!}{
m!}\right]$. Note that since $\lambda_{u}>0$ for $k>0$, the
function $u(t)$ associated with the dispersion will exponentially
increase in the process of evolution, while $v(t)$ related to the
absorption/amplification will decrease rapidly in this process. In
a decoupling case where $k=0$, $\lambda_{u}=0$. In this case, the
real part of polarization ({\it i.e.}, $u$) will not vary with the
development of the quantum system. However, the imaginary part
($v$) of polarization does not vanish.
\\ \\
\textbf{Acknowledgements} This project was supported by the
National Natural Science Foundation of China under Project No.
$90101024$.

\end{document}